\documentclass[journal]{IEEEtran}
\usepackage[linesnumbered,ruled,vlined]{algorithm2e}

\usepackage{color,soul,graphics,graphicx,amsmath,amsfonts,amssymb,times,setspace,cite,fancyhdr,epstopdf}
\usepackage{booktabs}
\usepackage{multirow}
\usepackage{float}

\usepackage{tikz}
\usepackage[ruled,vlined]{algorithm2e}
\usepackage{bm}
\usepackage{relsize}
\usepackage{pgfplots}
 \usetikzlibrary{external}
\usetikzlibrary{arrows,shapes, decorations.pathmorphing,backgrounds,positioning}

\DeclareMathSymbol{\N}{\mathbin}{AMSb}{"4E}
\DeclareMathSymbol{\Z}{\mathbin}{AMSb}{"5A}
\DeclareMathSymbol{\R}{\mathbin}{AMSb}{"52}
\DeclareMathSymbol{\Q}{\mathbin}{AMSb}{"51}
\DeclareMathSymbol{\I}{\mathbin}{AMSb}{"49}
\DeclareMathSymbol{\C}{\mathbin}{AMSb}{"43}
\DeclareMathSymbol{\Exp}{\mathbin}{AMSb}{"45}
\DeclareMathSymbol{\Prob}{\mathbin}{AMSb}{"50}
\usepackage{comment}
\newcommand{\norm}[1]{\lVert#1\rVert}
\usepackage[super]{nth}
\ifCLASSINFOpdf
\else
\fi
\ifCLASSOPTIONcompsoc
 \usepackage[caption=false,font=normalsize,labelfont=sf,textfont=sf]{subfig}
\else
 \usepackage[caption=false,font=footnotesize]{subfig}
\fi
\begin{document}
\bstctlcite{IEEEexample:BSTcontrol}
%
\title{Learning the Gap in the Day-Ahead and Real-Time Locational Marginal Prices in the Electricity Market   }
%
%
%

\author{Nika Nizharadze, 
         Arash Farokhi Soofi, \emph{Graduate Student Member, IEEE} and Saeed D. Manshadi,~\IEEEmembership{Member,~IEEE}
        \thanks{Arash Farokhi Soofi is with University of California San Diego, San Diego,
CA, 92161 USA. e-mail: afarokhi@ucsd.edu. Nika Nizharadze and S. D. Manshadi are with San Diego State University, San Diego,
CA, 92182 USA. e-mail: nnizharadze@sdsu.edu; smanshadi@sdsu.edu.}
}

\maketitle
{
\begin{abstract}
In this paper, statistical machine learning algorithms, as well as deep neural networks, are used to predict the values of the price gap between day-ahead and real-time electricity markets.  
Several exogenous features are collected and impacts of these features are examined to capture the best relations between the features and the target variable.
Ensemble learning algorithm namely the Random Forest is used to calculate the probability distribution of the predicted electricity prices for day-ahead and real-time markets. Long-Short-Term-Memory (LSTM) is utilized to capture long term dependencies in predicting direct gap values between mentioned markets and the benefits of directly predicting the gap price rather than subtracting the predictions of day-ahead and real-time markets are illustrated. Case studies are implemented on the California Independent System Operator (CAISO)’s electricity market data for a two years period. The proposed methods are evaluated and neural networks showed promising results in predicting the exact values of the gap.

\end{abstract}
}


%
\IEEEpeerreviewmaketitle

\vspace{-0.2cm}
\section{Introduction}

\IEEEPARstart{O}ne major concern in the design of a two-settlement electricity market is the gap in the clearing prices across day-ahead and real-time markets. Many electricity system operators have implemented a two-settlement market approach, which consists of Day-Ahead Market (RTM) and Real-Time Market (RTM), each producing its financial settlements. The DAM is financial market that schedules the production and consumption before the operating day, while the RTM is a physical market that settles based on the served demand and provided supply. The difference between the Locational Marginal Pricing (LMP) values of DAM and RTM is an indicator of the surplus or the shortage of electricity in the electric grid. Multiple settlements establish more stable prices and also reduces the vulnerabilities to price spikes in the real-time market~\cite{Veit2006}. It is shown in \cite{993956,Yao2008} that with two settlement electricity market, generation units have incentives to enter into real-time contracts, which will reduce real-time electricity prices, which in turn will increase social welfare. Consequently, both parties’ bidders, as well as generation companies, benefit from such settlement.
Nevertheless, there will be a gap between the day-ahead and real-time settlement. The increase in the penetration level of renewable energy resources exacerbated the volatility of energy supply and prices within the RTM and as result prediction the gap between DAM and RTM price become more complicated. Predicting such a gap plays an integral role in establishing the operating schedules and adjusting the bidding strategies of the market participants within the market. This is particularly important for convergence bidders who are interested in a hedge against the price gap across the two markets. CAISO operates day-ahead and real-time markets in which buyers and sellers can engage in convergence bidding activities. 
The Day-Ahead Market (DAM) is a financial market where participants submit their bids for the following 24 hours, whereas the Real-Time Market (RTM) is a physical market in which buyers and sellers submit their bids during the day usually for a 5-minutes interval. The RTM balances out the differences between DAM purchases and the actual real-time demand and generation of electricity. In this paper, the focus is on predicting the gap between the cleared prices within the two markets. The gap value can give significant insights about the supply and the demand for the electricity market which could be valuable information for both system operators as well as market designers and market participants to help them reduce their risks and enhance the efficiency of the market. 
\par
 
 

The prediction of day-ahead hourly electricity prices leveraging an integrated machine learning model is proposed in \cite{Fan2006}. In this article, authors employed Bayesian Clustering by Dynamics to cluster the data set into several subsets and Support Vector Machines (SVM) are used to fit the training data in each subset. The error metrics of the integrated model are significantly improved compared to that of the single SVM network. In~\cite{Contreras2003}, authors proposed Auto-Regressive Integrated Moving Average (ARIMA) models to predict next-day prices for Spanish and Californian markets. In~\cite{Sadeghi-Mobarakeh2017}, Random Forest regression is leveraged to predict DAM prices, the proposed approach outperformed the ARIMA model, however, this paper does not consider the impacts of other features such as temperature, and solar radiance to predict prices or price gap.

Even though statistical models perform well at identifying patterns and indicators that will influence the price of the asset, they struggle to predict prices accurately in the presence of spikes which is particularly important for predicting the gap price across a two-settlement market ~\cite{Weron2014}. The electricity market depends on many characteristics such as weather, temperature, wind speed, and precipitation. Thus, the LMP tends to fluctuate over an operation horizon. Consequently, to handle the complexity of the power market, Artificial Neural Networks (ANNs) are used in~\cite{Adebiyi2014}.
The increase in the number of computation layers increases the feature abstraction capability of the network, which makes them better to identify non-linear trends~\cite{Bengio2013}. In~\cite{Zheng2017}, the LSTM network and a variation of deep Recurrent Neural Network (RNN) are used to forecast electricity load and the outputs of the model are compared to that of statistical models. Electricity consumption of the past 10 days is used to predict the electricity consumption of the next day. The LSTM-based model significantly outperformed the Seasonal-ARIMA and Support Vector Regression (SVR) models. A similar model is used in~\cite{Bouktif2018} to predict electricity load, but in this case, in addition to the historical load data, weather datasets are also utilized, however no change in the model performance is observed when weather-related features are removed and only the time lags are used as inputs. In~\cite{Jiang2018} the LSTM network is used to predict future 24 hours of prices for electricity prices for Australian and Singaporean markets. The MAPE is used to evaluate the model and up to 47.3\% improvement was observed compared to Multi-layer ANN. 

According to~\cite{Aggarwal2009}, the prediction of real-time LMP is even more challenging and most of the approaches adopted from previous studies generate mean absolute percentage error (MAPE) around 10-20\%. In~\cite{Ji2013}, homogeneous Markov chain representation of RTM LMP is used to predict RTM LMP for the prediction horizon of 6-8 hours. Future prices are computed based on state transition matrices, using the Monte Carlo method. Although the MAE metric of the model was 11.75, it has a huge computation burden.
Authors of the paper~\cite{Mujeeb2019}, proposed deep LSTM (D-LSTM) to predict short and medium-term load and LMP.
The D-LSTM turned out to be a flat trend without the validation set, however, once the network is tuned, it outperformed the Nonlinear Auto-regressive network with Exogenous variables (NARX) and Extreme Learning Machine (ELM) models in terms of accuracy. In~\cite{Zhang2020}, researchers used the Generative Adversarial Network (GAN) based video prediction approach on market data from ISO NE to predict RTM LMPs. The market data images are created from the historical data and by concatenation of these images, a video stream is created. Consequently, the prediction of the next frame is used to predict the next-hour RTM LMP. The proposed method achieved approximately 11\% MAPE score, however, weather data sets are not utilized to better model the price spikes. The enhanced convolutional neural networks are also used in~\cite{Zahid2019}, to predict electricity load and prices. Here, feature selection is done using the Random Forest model, and extracted features are passed to the convolutional layer, which later is filtered using the Max Pooling layer. Showcased work resulted in smaller error measurements than SVR using NYISO market data. 

Leveraging the LSTM to predict the gap between RTM and DAM prices using weather features brings the following questions to the mind: \textit{What are the advantages employing to predict gap of prices across RTM and DAM compared to other approaches?} \textit{Can we improve the prediction of price gaps by leveraging exogenous information (e.g. weather data including solar irradiance)?}

 The contributions of this paper are summarized as follows:
  \begin{enumerate}
\item An extensive dataset was collected, which includes information about the electricity market as well as the coordinated weather data. The impacts of using an external dataset are illustrated and the significance of features is shown.
\item Both DAM and RTM are analyzed for price prediction and the correlation between the features from these markets is demonstrated. Furthermore, a realistic set of assumptions is made regarding the availability of features for each of RTM and DAM once the prices are predicted for the following market operation day upon clearing of the market.
\item Ensemble learning method namely the Random Forest (RF) is used to calculate the probability distribution of the predicted market prices for the DAM and RTM.
\item Long Short Term Memory (LSTM) architecture is deployed to handle the complexity of the data set. The proposed model is compared to statistical machine learning methods and significant improvements are observed.
%
 \end{enumerate}

 
\vspace{-0.2cm}
\section{Learning Algorithms and Methodologies}\label{Learning_methods}
In this Section, the methods which are leveraged to examine the direct price gap values are introduced. Described learning algorithms are utilized to predict price gaps as well as to rank features based on their importance and to construct probability distributions for DAM and RTM price predictions.
\vspace{-0.2cm}
\subsection{Least Absolute Shrinkage and Selection Operator (LASSO)}
The objective of the Linear Regression model is to find a relationship between two variables by fitting a linear equation to observe data points. The most common way to find a fitted line is to use the Least-Squares method in which the model finds a fitted line by minimizing the sum of squared residuals, however by shrinking or setting some coefficients to 0 can increase the accuracy of the mentioned model~\cite{Tishbirani1996}. In the Lasso model, $L_1$ regularization term is added to the cost function to address the above-mentioned issue. The penalty term $\lambda\Vert\theta\Vert_1$ is the absolute value of the magnitude of the parameters, where $\lambda$ controls the amount of regularization. Large enough values of $\lambda$ will cause shrinkage and set the regression coefficients exactly equal to 0. After the shrinkage occurs, only the non-zero parameters are taken into consideration. Shrinking the regression coefficients reduces variance and minimizes the bias. However, the accuracy of the model heavily relies on the parameter $\lambda$ which controls the factor of shrinkage. When $\lambda$ becomes 0, the model performs exactly the same as Linear Regression model. When $\lambda$ increases the variance is significantly decreased. Lasso is good method to eliminate the irrelevant variables and only consider related variables to compute the output of the model.
The cost function $J$ of the LASSO method is presented in (\ref{eq:LASSO_cost_function}).
\vspace{-0.2cm}
\begin{equation}\label{eq:LASSO_cost_function}
    J^{}_{_{\text{LASSO}}}(\theta) = \frac{1}{2}\sum_{i=1}^{m}(g^{}_{_{\text{LASSO}}}({{\mathbf{x}}}_i) - y_i)^2 + \lambda\Vert\theta\Vert_1
\end{equation}
\vspace{-0.6cm}
 \subsection{Support Vector Regression (SVR)}\label{method:svr}
The SVR method is a non-linear learning algorithm. One of the most common versions of SVR regression is $\epsilon$-SV regression. The goal of $\epsilon$-SV regression is to find a function that has at most $\epsilon$ divergence for all the data points. The algorithm accepts the errors only within the range of  $\epsilon$ as presented in \eqref{eq:SVR_constraint}. 
\vspace{-0.2cm}
\begin{subequations}
\begin{alignat}{2}
&\!\min_{{{\mathbf{w}}},b,\xi,\xi^*}        &\qquad& \frac{1}{2}\norm{{{\mathbf{w}}}}_2^2 + C\sum_{i=1}^{m}(\xi_i+\xi_i^*) \label{eq:SVR_optProb}\\
&\text{subject to} &      & {{\mathbf{w}}}^T\phi({{\mathbf{x}}}_i) + b - y_i \leq \epsilon + \xi_i\label{eq:SVR_constraint1}\\
&                  &      & y_i - {{\mathbf{w}}}^T\phi({{\mathbf{x}}}_i) - b  \leq \epsilon + \xi_i^*\label{eq:SVR_constraint2}\\
&                  &      & \xi_i, \xi_i^* \geq 0, i = 1,\cdots,m\label{eq:SVR_constraint3}
\end{alignat} \label{eq:SVR_constraint}
\end{subequations}
Here, constant $C > 0$ balances the flatness of a function, and the amount up to which deviations larger than $\epsilon$ are tolerated. $\phi({{\mathbf{x}}}_i)$ maps ${{\mathbf{x}}}_i$ into a higher-dimensional space, where ${{\mathbf{w}}}$ and $b$ are coefficients. $\xi$ and $\xi^*$ represent the distance from the actual values to the margin of the $\epsilon$-tube with support vectors. Errors outside the margin are penalized linearly. 

The dual formulation of the $\epsilon$-SVR method provides the key for extending SV machine to nonlinear functions. Hence, a standard dualization method utilizing Lagrange multipliers is leveraged as presented in \eqref{eq:SVR_dual_constraint}.
\begin{subequations}
\begin{alignat}{2}
&\!\min_{\alpha_i, \alpha_i^*}        &\qquad& \frac{1}{2}(\boldsymbol{\alpha}-\boldsymbol{\alpha^*})^TQ(\boldsymbol{\alpha}-\boldsymbol{\alpha^*})\nonumber\\
&                  &      & +\epsilon\sum_{i=1}^{m}(\alpha_i+\alpha_i^*) +\sum_{i=1}^{m}y_i(\alpha_i-\alpha_i^*)\label{eq:SVR_dual_optProb}\\
&\text{subject to} &      & {\mathbf{e}}^T(\boldsymbol{\alpha}-\boldsymbol{\alpha^*})=0,\label{eq:SVR_dual_constraint1}\\
&                  &      & 0\leq \alpha_i, \alpha_i^* \leq C, i = 1, \cdots, m\label{eq:SVR_dual_constraint2}
\end{alignat} \label{eq:SVR_dual_constraint}
\end{subequations}
In the SVR model the objective is to fit the error within an $\epsilon$ threshold. To achieve this, the algorithm constructs a decision boundary which has $\epsilon$ distance from the original hyper-plane such that support vectors or closest data points to the hyper-plane are within the decision boundary line. Only the points within the decision boundary are taken into consideration, consequently only the points with least errors are used, thus the model is better fitted in comparison to LASSO method. 

Here, $\alpha_i, \alpha_i^*$ are Lagrange multipliers, ${\mathbf{e}}^T$ is the all ones vector, and $Q$ is an $m \times m$ positive semi-definite matrix. $Q_{k,l}=K({{\mathbf{x}}}_k,{{\mathbf{x}}}_l)$, where $K({{\mathbf{x}}}_k,{{\mathbf{x}}}_l)$ is the radial basis kernel function (RBF), so $K({{\mathbf{x}}}_k,{{\mathbf{x}}}_l) = \text{exp}(-\gamma\norm{{{\mathbf{x}}}_k-{{\mathbf{x}}}_l}^2)$, where $\gamma$ is kernel parameter.
{Once the dual problem is solved and Lagrange multipliers are determined, the optimal weights and base can be computed. A predictor $g$ of SVR with $m$-training examples is presented in \eqref{eq:SVR_function_presentation}.
\vspace{-0.15cm}
\begin{equation} \label{eq:SVR_function_presentation}
g^{}_{_{\text{SVR}}}({{\mathbf{x}}})=\sum_{i=1}^{m}(-\alpha_i+\alpha_i^*)K({{\mathbf{x}}}_i,{{\mathbf{x}}})+b.
\end{equation} 
}

\vspace{-0.95cm}

\subsection{Random Forest Algorithm}\label{RandomForest}
Random Forest is an ensemble learning algorithm. It combines multiple weak models to build a strong predictor by taking advantage of methods called Bagging and Decision Trees. The goal of the Decision Tree algorithm is to build a tree-like structure from the existing data points, where each leaf will only contain labels from the same class. The algorithm will split the dataset into roughly two halves until the leaves are pure. 

To find the best split which will keep the tree compact, the impurity function is minimized. In the case of regression tasks usually Squared Loss as given in ~\eqref{SquaredLoss} is used as an impurity function, while classification problems employ Gini impurity as presented in ~\eqref{Gini}. 
\vspace{-0.10cm}
\begin{subequations}\label{SquaredLoss}
\begin{align}
   L(D)=\frac{1}{|D|}\sum_{(x,y)\in D}(y-\bar{y}_D)^2
    \label{rf1a}\\
    \textrm{where }\bar{y}_D=\frac{1}{|D|}\sum_{(x,y)\in D}y
    \label{rf1b}
\end{align}
\end{subequations}
\vspace{-0.05cm}

Given a dataset $D=\left \{ \left ( \mathbf{x}_1,y_1 \right ),\dots,\left ( \mathbf{x}_n,y_n \right ) \right \}$ with $c$ number of distinct categories where $D_k$ is all inputs with label $k$, squared loss impurity outputs average squared difference of actual value and average prediction, while Gini impurity measures homogeneity of classes. 
\vspace{-0.15cm}
\begin{subequations}\label{Gini}
\begin{align}
    G(D)=\sum_{k=1}^{c}p_k(1-p_k)
    \label{rf2a}\\
    \textrm{where } p_k=\frac{\left | D_k \right |}{\left | D \right |}
    \label{rf2b}
\end{align}
\end{subequations}
Decision Trees are learning the exact patterns in the training set, thus they don't generalize well enough, so they are prone to overfitting. Random Forest is using Bagging to decrease the high variance caused by Decision Trees. Bagging generates $D_1,...,D_m$ datasets from the existing data points $D$. The created datasets are the replicated datasets each consisting of $k$ features drawn at random but with replacement from the original dataset~\cite{Breiman1996}. The new datasets are equal in size to the original dataset and have approximately the same probability distribution. \par
Random Forest consists of large number of Decision Trees ${h(x,D_m)}$ from $D_1,...,D_m$ where $D_m$ is an independent identically distributed vector~\cite{Breiman2001}. In the case of classification tasks the majority vote acquired form all the Decision Trees will be the prediction, and for regression purposes, the average of all predictors will be the output. Moreover, the Random Forest Algorithm has only two hyper-parameters $m$ and $k$. Based on empirical evidence good choice for $k$ is $k=\sqrt{d}$, where $d$ is the total number of features in the dataset, and increasing the size of $m$ will only benefit the model.


\subsection{Long-Short Term Memory (LSTM)}
Neural Networks try to model the behavior of the human brain. They consist of artificially created neurons and a set of edges that connect those neurons. Besides, each neuron has its associated activation function, which models neuron impulses. Recurrent Neural Network (RNN) is a special type of neural network where the input is the sequence. RNN is very powerful because it not only uses the input to predict the output but it also utilizes the information from previously observed timestamps. All RNNs form a sequence of connected units which represent the state of the network at timestamp $t$. Single module takes data from previous unit $h_t-1$ and input for that timestamp $x_t$, then it computes output for timestamp $t$ using $\tanh$ function. According to~\cite{Sontag1991}, a finite-sized recurrent neural network can compute any function that exists. However, RNNs suffer either from exploding or vanishing gradients when back-propagating through time. To update weights, the neural network is computing partial derivatives of the loss function of the current layer at each timestamp. Consequently, when gradients are very small either learning happens at a very slow rate or does not happen at all. To overcome this issue with RNNs, the Long-Short Term Memory (LSTM) model is proposed as suggested in~\cite{Hochreiter1997}. LSTM is a special kind of recurrent neural network architecture. Instead of only using $\tanh$ function in a unit, LSTM utilizes three gate units: forget gate, input gate, and output gate. 
The forget gate is responsible to keep only the relevant information as given in~\eqref{lstm1}. It takes input at timestamp $x_t$ and the data from previous hidden layer $h_t-1$, then the $sigmoid$ function is applied to those inputs, thus as a result the output of the forget gate is in between 0 and 1. The output closer to 0 will be forgotten, and output with the numeric value of 1 will be kept for further calculations. Besides, the input gate decides how the memory cell will be updated as shown in~\eqref{lstm2}. First, the candidate value is computed using~\eqref{lstm3}, then the result is scaled by the output of the input gate to decide by how much the cell state will be updated. Finally, LSTM employs an output gate which is a filtered version of the cell state. At first, the cell state is normalized using $\tanh$, then the $sigmoid$ layer is utilized to decide which parts of the memory will be output. The output of the hidden state $h_t$ and the prediction  $y_t$ is the same, however notation $h_t$ is used as an hidden state input at timestamp $t+1$. The structure of the LSTM network for a single unit is given in Fig. 1.
\vspace{-0.2cm}
\begin{subequations}\label{lstm}
\begin{alignat}{2}
    f_t &= \sigma(W_{f}[h_{t-1}, x_t] + b_f)
    \label{lstm1}\\
    i_t &= \sigma(W_{i}[h_{t-1}, x_t] + b_i)
    \label{lstm2}\\
    \tilde{c_t} &= \tanh(W_{c}[h_{t-1}, x_t] + b_c)
    \label{lstm3}\\
     c_t &= f_t \otimes c_{t-1} + i_t \otimes \tilde{c_t}
    \label{lstm4}\\
    o_t &= \sigma(W_{o}[h_{t-1}, x_t] + b_o)
    \label{lstm5}\\
    h_t &= \tanh(c_t) \otimes o_t
    \label{lstm6}
\end{alignat}
\end{subequations}
\vspace{-0.8cm}
\begin{figure}[!h] 
\centering
\includegraphics[width=0.5\textwidth]{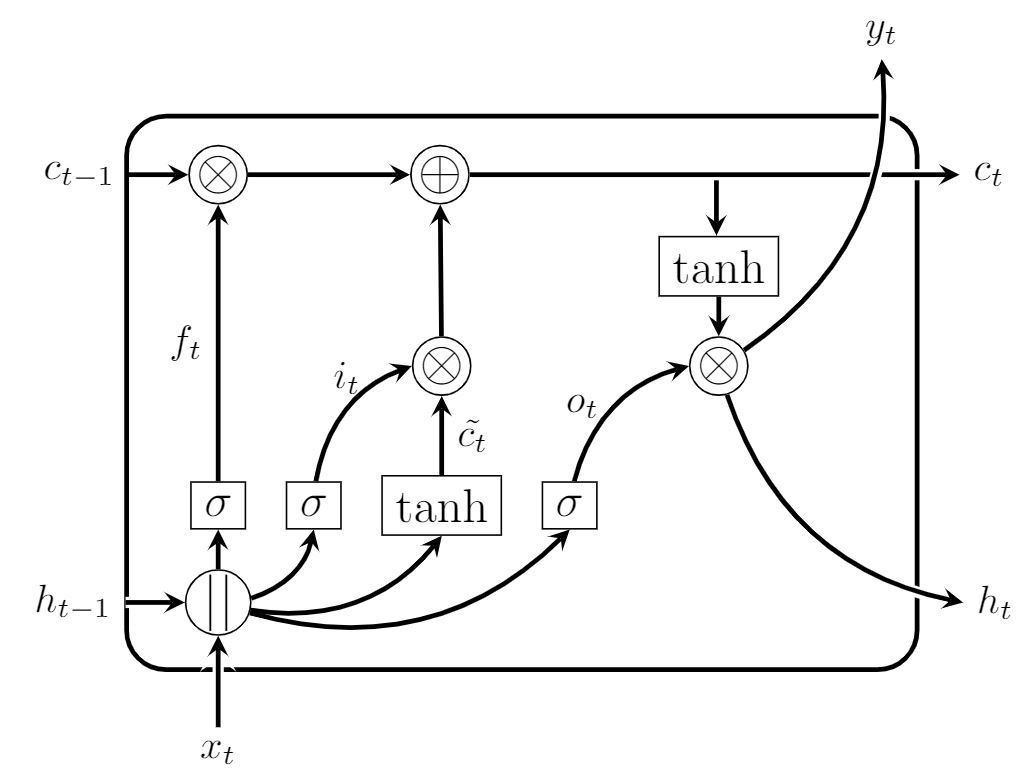}
\caption{The LSTM architecture of a single unit.} \end{figure}
\vspace{-0.51cm}
\section{Prediction Performance Evaluation}

\begin{table}[h!] 
\small \centering
\centering
\vspace{-0.21cm}
\caption{Prediction errors for Station A while using temporal data for various learning algorithms}\vspace{-0.2cm}
\begin{tabular}{ll} \hline \hline
\textbf{MAE} &
\begin{math}
    \left[\frac{1}{s}\sum_{t=1}^{s}|y_t - \hat{y}_t|\right]
\end{math} \\
\hline 
\textbf{RMSE} &
\begin{math}
    {\sqrt{\frac{1}{s}\sum_{t=1}^{s}(y_t - \hat{y}_t)^2}_{}}
\end{math} \\ \hline 

\textbf{nRMSE} &
\begin{math}
    {\sqrt{\frac{1}{s}\sum_{t=1}^{s}(y_t - \hat{y}_t)^2}_{ }\bigg/
    \left[y_{\text{max}}-y_{\text{min}}\right]}
\end{math} \\ \hline 
\textbf{Max Error} &
\begin{math}
\label{eqn:max_error}
    max(\left|{y_t - \hat{y}_t} \right|) \forall t \in \{0, s\} 
\end{math} 

\\ \hline \hline

\end{tabular} \label{all_errors}
\vspace{-0.21cm}

\end{table}



The metrics introduced in Table \ref{all_errors} are presented to measure the performance of learning algorithms presented in the previous section. Here, $y_t$ is the actual value at time $t$, while $\hat{y}_t$ is the predicted value for the same timestamp. $y_{\text{max}}$ and $y_{\text{min}}$ represent maximum and minimum values of all actual values. In addition, $s$ is the number of samples in the testing dataset. Mean-absolute error (MAE) described in Table \ref{all_errors} measures the average magnitude of the errors between predictions and actual values. Similarly, RMSE represented in Table \ref{all_errors} also expresses average model prediction error, however, it is measured by taking a square root from the average of squared differences between the actual and predicted values. Both of the mentioned metrics measure prediction errors and they can range from 0 to $\infty$. Consequently, lower values characterize the better performing model. 
Error metric nRMSE outlined in Table \ref{all_errors} represents the normalised RMSE value. In this case, the normalization is done by dividing the RMSE score by the difference of the maximum and the minimum values of actual values. Furthermore, the metric max error described in Table \ref{all_errors} represents the evaluation of the worst-case scenario and it measures the maximum difference between the predicted value and the ground truth.

\section{ Data Preparation}

\subsection{Data Collection}
Three distinct datasets are collected and merged to form an extensive dataset for studying the gap between DAM and RTM.
The California Independent System Operator (CAISO) provides an Open Access Same-time Information System (OASIS) Application Programming Interface (API), which produces reports for the energy market and power grid information in real-time. To demo the results of this paper, the \textit{MURRAY6N015} node, located in San Diego, CA, is chosen and its reports for the energy market and power grid information in real-time are leveraged.
The LMP, the LMP congestion component, the LMP energy component, and the LMP loss component are collected for DAM and RTM. Besides, a seven-day ahead load forecast as well as load forecasts for the next two days are acquired using the above-mentioned API. The period of the collected dataset is two years starting from the \nth{1} of January 2017. 
\par
For the historical hourly weather dataset, Meteostat API is used. Meteostat is collecting hourly weather measurements from more than 5000 weather stations around the world. Besides, they offer comprehensive historical datasets that combine their measurements with the NOAA's Global Historical Climatology Network's dataset. Weather data is obtained from the San Diego International Airport weather station, which is the closest weather station to the node of interest. Collected data includes information about temperature, dew point, humidity, wind speed, wind direction, weather condition, sea level pressure, wind gust, cloud layers, and weather forecasts for the next 3 and 6 days.
\par
Besides, mesonet API is utilized to acquire a dataset for solar radiation. Mentioned API offers quality controlled, surface-based environmental data such as Global Horizontal Irradiance (GHI), Direct Normal Irradiance (DNI), Diffuse Horizontal Irradiance (DHI), solar zenith angle, cloud type, and precipitable water. GHI is the total amount of terrestrial irradiance received from above by a surface horizontal to the ground. DNI means the radiation that comes in a straight line directly from the sun and is absorbed by a unit perpendicular to the rays. Furthermore, DHI is the radiation that does not arrive on a direct path from the sun, and it is equally absorbed from the particles in the atmosphere. The San Diego International Airport weather station is picked to collect the above-mentioned measurements.

\vspace{-0.1cm}
\subsection{Data Cleansing and Pre-Processing}
The above-mentioned datasets are merged based on the date and the hour of the day. Only the data from the time span of January \nth{1} 2017, 00:00 to December \nth{30} 2018, 23:00 is utilized. Data cleansing techniques are applied to ensure the quality of the data. Duplicate rows are dropped, categorical variables are converted to numerical representations, and every measurement is converted to a floating-point value. Also, the missing values are substituted with a global constant. After data cleansing, $16566$ hours worth of data is available. The dataset is arbitrarily split into two parts. The $90\%$ of the data is used for training and the remaining $10\%$ is utilized for testing purposes.

Then, the input data is normalized using Min-Max scaler thus every feature is converted in a range of \{0,1\}. The Random Forest model is employed to select Feature. The data collected is extensive and combines three different datasets, consequently, it is important to showcase which features contribute to prediction and which are insignificant. Moreover, feature selection ensures that features that do not affect the prediction are removed and do not introduce extra noise in the system.

\section{Simulation Results}



\subsection{Feature Importance}

Lasso, SVR, and Random Forest algorithms can not inherently capture temporal dependencies for sequential data, that's why day-ahead prices for a previous 48-hour time horizon are added as features. Therefore, 48 new columns are created which contain the delayed values of DAM LMP. Similarly, lagged real-time and gap values are added to the existing dataset, however for the RTM, at any given time $t$ the most recent prices that are available are at $t-12$ hour. Consequently, only those features that are realistically available for RTM are taken into consideration. \par
Since Random Forest method employs decision trees, it is a good approach to leverage Random Forest method for feature selection. Random Forest naturally ranks by how well each decision tree is improving the purity of the node. The Gini index of decision tree algorithms is leveraged as feature importance values. For example, the greatest decrease in impurity happen at the root of the tree, while the least decrease in impurity happens at the leaves of the tree. Consequently, by pruning the tree below a particular node creates a subset of the most relevant features.
Table~\ref{temporal_A} presents features ranked by their importance for gap predictions. The mentioned algorithm showed that 204 features are useful for gap prediction. Note that 14 most relevant features are shown in Table~\ref{temporal_A}.
The right column of the Table~\ref{temporal_A} represents the importance coefficient. The importance coefficient is scaled so that sum of all importance coefficients is 100. The most informative feature is DAM LMP price in the previous hour, which is reasonable because the most recent price has to have the most influence on price change for the next hour. In case of RTM for each hour there are 12 LMP price points available and each 5 minute interval is numbered from 1 to 12. Consequently, \textit{RTM 1 LMP price 24 hours ago} corresponds to the RTM LMP price at 23 hours and 55 minutes ago. The importance coefficient of the above-mentioned feature is 1.3. Besides, the external features that are collected demonstrated a significant effect on predicting the price gap between the two markets. For example, the solar zenith angle has an importance coefficient of 1.0, while DHI contributes to prediction with 0.39 importance score. 
\par

\begin{table}[h!]
\small \centering
\centering
\vspace{-0.21cm}
\caption{Feature Importance for Gap predictions.}\vspace{-0.2cm}
\begin{tabular}{|c|c|} \hline 
\textbf{Feature} & \textbf{Importance Coefficient}\\ \hline 
DAM LMP price 1 hour ago  & 9.18   \\ \hline 
GAP LMP price 24 hours ago  & 3.32  \\ \hline 
DAM LMP price 24 hours ago  & 1.9   \\ \hline 
RTM 1 LMP price 24 hours ago  & 1.3   \\\hline 
Solar Zenith Angle  & 1.0   \\\hline 
Demand Forecast Day-Ahead  & 0.78   \\\hline 
Relative humidity  & 0.67   \\\hline 
Perceptible Water  & 0.61   \\\hline 
Cloud Layer  & 0.54   \\\hline 
Dewpoint  & 0.53   \\\hline 
Wind Speed  & 0.47  \\\hline 
Wind Direction  & 0.46   \\\hline 
Demand Forecast 2 Days Ahead  & 0.46   \\\hline 
DHI  & 0.39   \\\hline 
\end{tabular}
\vspace{-0.21cm}
\label{temporal_A}
\end{table}
\vspace{-0.51cm}
\subsection{Hyper-parameter Tuning}
To perform day-ahead, real-time, and gap predictions, the hyper-parameters of each learning methods presented in Section \ref{Learning_methods}, are optimized. Hyper-parameters control the learning process and they have to be optimized such that the predefined loss function is minimized for a given dataset. Grid search with cross-validation is used to tune hyper-parameters. Grid search is a brute force algorithm, which calculates the output for all subsets of predefined parameters and picks the best estimator. The performance of the estimators is evaluated using K-Fold cross-validation, where $k=5$. K-Fold cross-validation divides the existing dataset randomly into $k$ groups of data and fits the model using $k-1$ groups of data as a training set while testing the model on the $k$-th fold. The process is repeated until every $k$ group serves as a testing set. \par
The Lasso algorithm presented in (\ref{eq:LASSO_cost_function}) has only one hyper-parameter $\lambda$. The set of arbitrarily chosen values  \{0.0001, 0.0002, 0.0003, 0.0004, 0.0005, 0.001, 0.002, 0.003, 0.004, 0.005, 0.01\} are examined to find the optimal value for $\lambda$. The highest accuracy or minimal loss is acquired using grid search when  $\lambda$ hyper-parameter is $0.0003$. \par
The SVR model is optimized for 4 different hyper-parameters including $C$,  $\epsilon$, kernel function $K$, and kernel coefficient $\gamma$. The optimal value found using grid search for constant $C$ is $C=1000$, while the margin of tube $\epsilon = 0.001$ turned out to give the most accurate estimator. In addition, the different kernel functions including: linear, sigmoid, and radial basis functions are tested and the most accurate results are obtained using radial basis kernel function with coefficient $\gamma = 0.1$ .\par
Similarly, the random forest algorithm is also tuned for hyper-parameters. Generally, the increase in the number of trees in the forest can only benefit the algorithm, however, the mentioned increment also introduces significant overhead in computation time, so the only forest with 50 and 100 trees is tested and 100 trees turned out to give more accurate results.
In addition, the maximum number of features considered when looking for the best split turned out to be equal to the number of all features. Moreover, different maximum depths of the trees are passed to grid search and the optimal value is found when the nodes are expanded until all leaves are pure. Finally, the algorithm is tuned for the methods of sampling the data points and sampling with replacement turned out to be a more optimal option.

LSTM model is optimized for the number of units, loss function, optimizer, and look back period which represents several previous timestamps that are considered for prediction at each time unit. The following set of values \{10, 20, 50, 100\} are examined for the number of units, while MAE and MSE are tested for loss function and Stochastic Gradient Descent (SGD), RMSProp, and Adam are utilized for optimizer choices. Note that RMSProp is a gradient based optimization technique that uses moving average of squared gradients to normalize the gradient, while Adam is a combination of RMSProp and SGD. Moreover, for lookback options a day, a week and a month are tested and for epoch number, the following set of values are examined \{10,20,50,100\}. It turned out that 100 LSTM cells with loss function of MSE and with adam optimizer eventuated into the most accurate results. In addition, the optimal lookback period is a day and the best number of epochs is 100.

 \subsection{Analysis of probability distributions}
\begin{figure}[t!]
\centering
\includegraphics[width=0.5\textwidth]{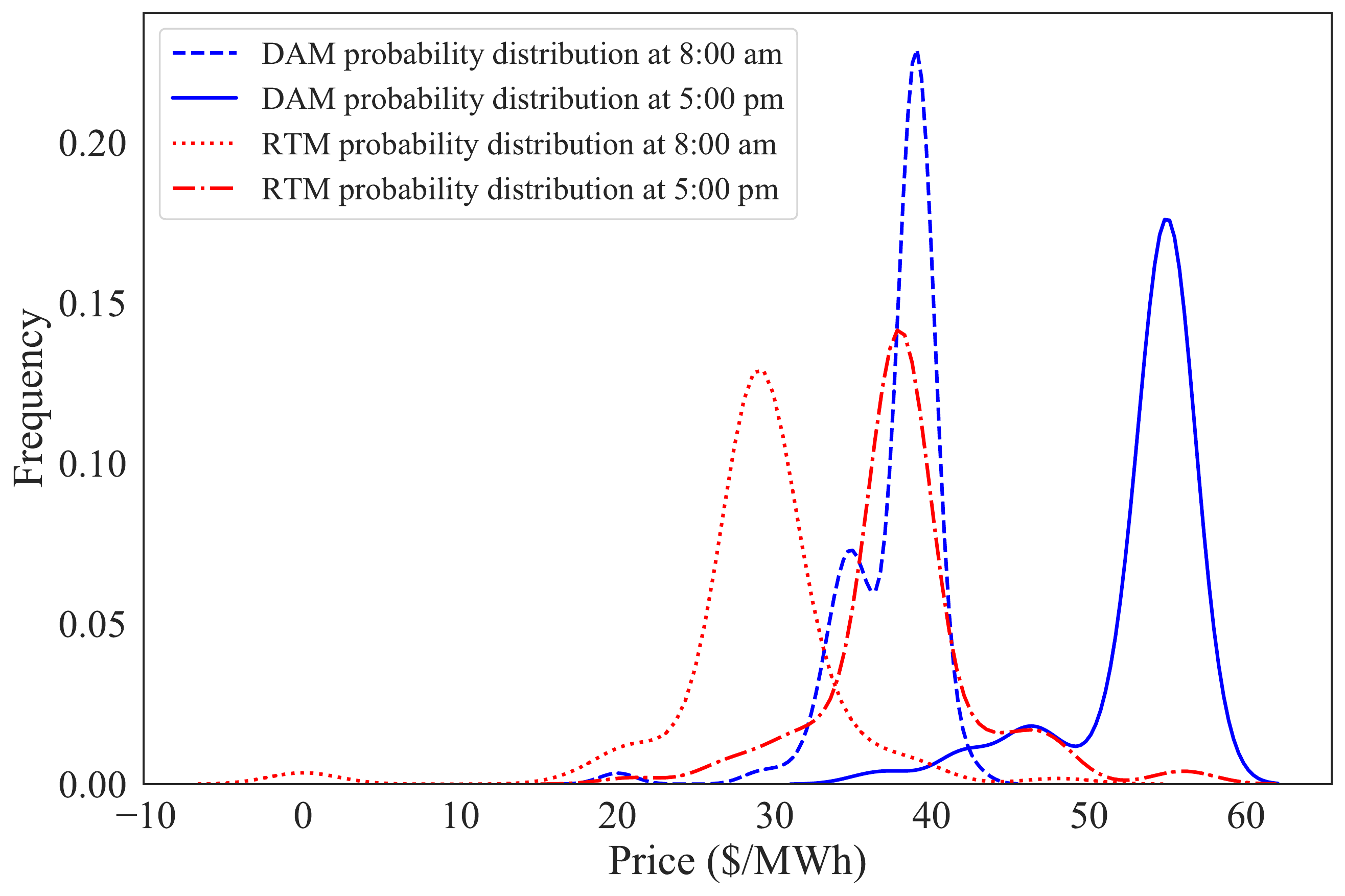}
\caption{The probability distribution of DAM and RTM price predictions for a specific date and time}
\label{fig:dam_rtm_distribution}
\end{figure}
\par

Random Forest algorithm described in Section \ref{Learning_methods} is used to calculate the probability distribution of the predicted electricity prices for the DAM and RTM. The outputs of the 100 regression trees are used to approximate the probability distribution for both markets. The prediction of each regression tree represents $1\%$ probability that the prediction is correct. To represent the results, we arbitrarily chose October \nth{15} 2018 and the hours of interests were 8:00 am and 5:00 pm. Fig.~\ref{fig:dam_rtm_distribution} shows the probability distributions for both of the markets at mentioned times. 
For the given day the prices for the DAM at 8:00 am range from 18 \$/MWh to 44 \$/MWh, while prices for the same market at 5 pm range from 30 \$/MWh to 62 \$/MWh. For this case study, the RTM prices tend to be in a lower range. Price prediction for RTM at 8 pm is in the range from -3 \$/MWh to 44 \$/MWh, while price prediction at 5 pm ranges from 17 \$/MWh to 60 \$/MWh. The electricity prices tend to be much higher at 5 pm compared to that of at 8 am, which is reasonable. 

\begin{figure}[t!]
\centering
\includegraphics[width=0.5\textwidth]{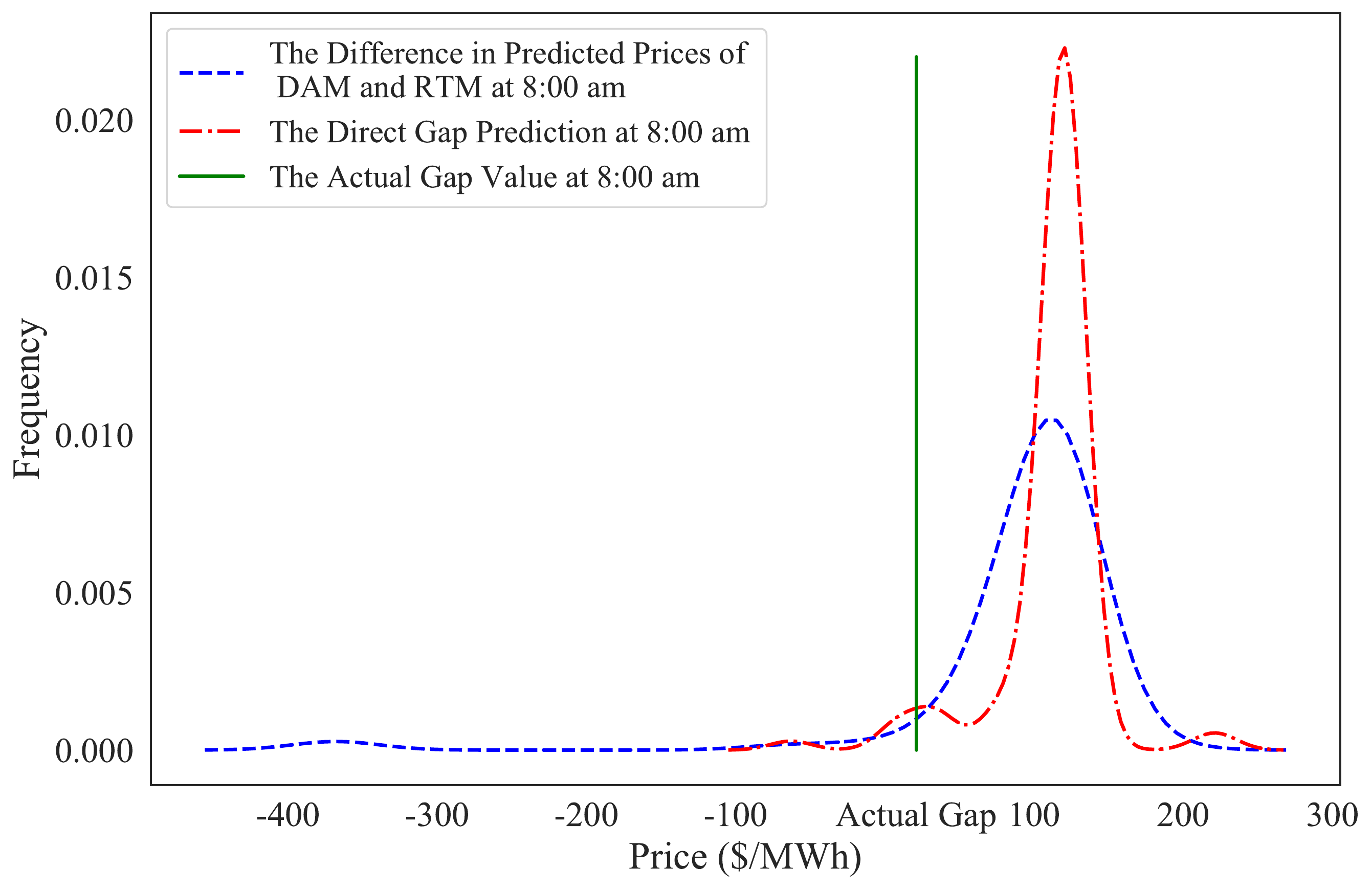}
\caption{The comparison between the probability distribution of direct gap prediction and the difference of separately predicted prices for DAM and RTM at 8 am}
\label{fig:gap_8am_distribution}
\end{figure}

\begin{figure}[t!]
\centering
\includegraphics[width=0.5\textwidth]{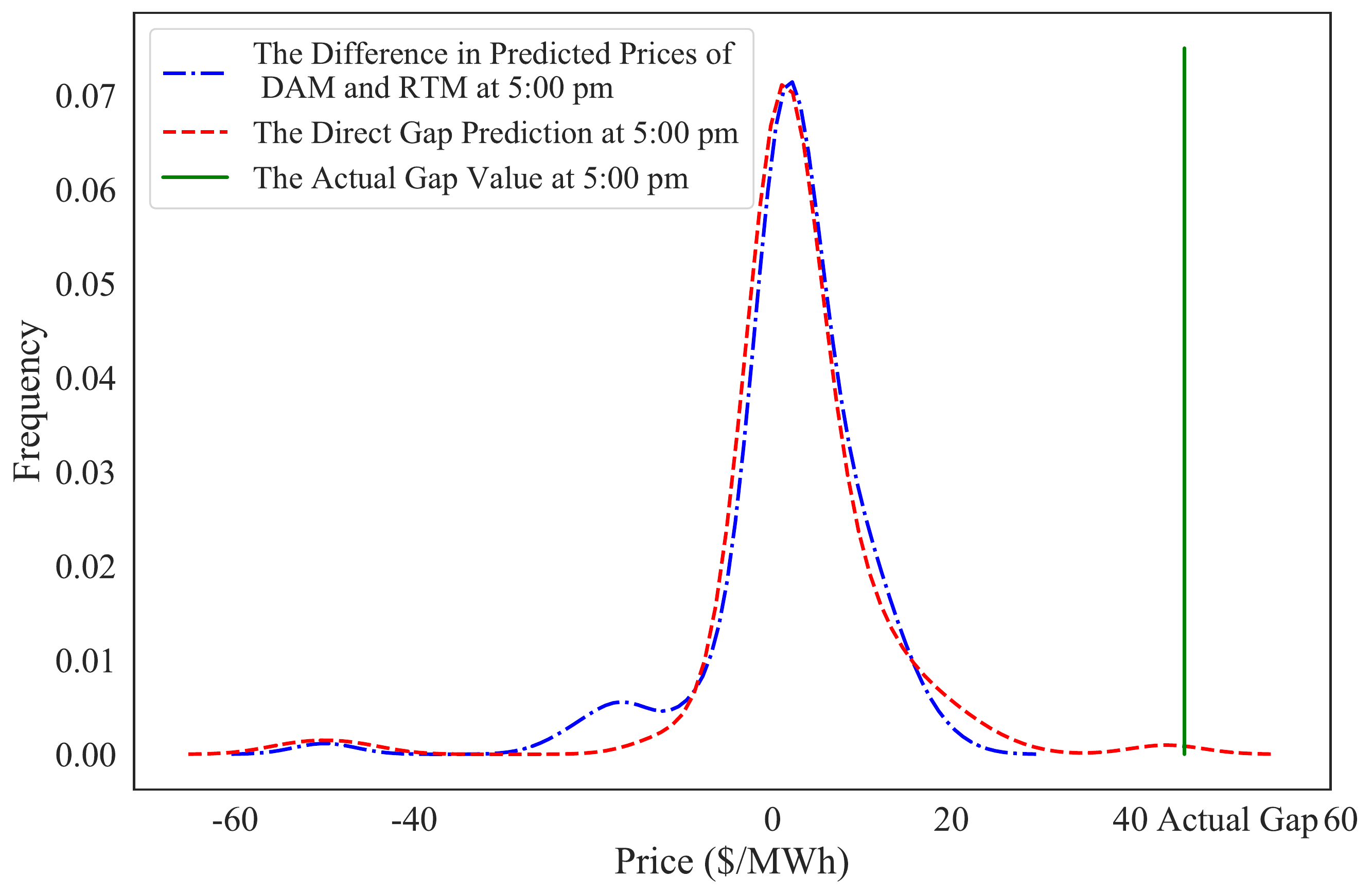}
\caption{The comparison between the probability distribution of direct gap predictions and the difference of separately predicted prices for DAM and RTM at 5 pm}
\label{fig:gap_5pm_distribution}
\end{figure}

Fig.~\ref{fig:gap_8am_distribution} and Fig.~\ref{fig:gap_5pm_distribution} present the importance of direct gap prediction in comparison to the difference in predicted prices of DAM and RTM. In direct gap prediction, the target value for the model is the gap price between the DAM and RTM, however to calculate the difference in predicted DAM and RTM, two distinct models are developed to predict DAM and RTM prices and then the predictions are subtracted. 
The time and date are the same as in the case study described above, however, in this case, the actual electricity gap price is also displayed to underline the significance of direct gap prediction. The blue dashed line represents the probability distribution acquired by subtracting predictions for DAM and RTM, while the red-dotted line is denoted for the probability distribution of the direct gap prediction, and the solid green line indicates the actual gap price at that specific hour. The actual gap price for the mentioned date at 8 am is 17.4 \$/MWh. 

It can be clearly observed from the Fig.~\ref{fig:gap_8am_distribution} that the probability of the gap to be 17.4 \$/MWh is higher in case of direct gap prediction rather than the probability acquired by subtracting the day-ahead and real-time price predictions. The probability distributions for 5 pm is presented in  Fig.~\ref{fig:gap_5pm_distribution}. It is obvious that direct gap prediction has more chances to be accurate. The actual gap price for 5 pm is  49 \$/MWh, however, the price range acquired by subtracting DAM and RTM price predictions is from -60 \$/MWh to 30 \$/MWh, consequently in this case it would be impossible to correctly predict the actual gap by calculating difference of predictions for mentioned markets. On the other hand, the range for direct gap predictions includes the actual gap value. Even though, in this case the probability of accurately predicting the actual gap using the direct gap prediction is not very high, it is still a better choice between those two methods.

\subsection{Performance Evaluation}
To understand the quantitative insights for the DAM, RTM, and gap prices, the descriptive statistics of the whole data set are presented in Table~\ref{tab:statistics}.
\begin{table}[h!]
\small \centering
\centering
\vspace{-0.21cm}
\caption{Statistical analysis of DAM, RTM, and gap prices of the data set}\vspace{-0.2cm}
\begin{tabular}{llll} \hline 
\textbf{Statistics} & \textbf{Gap} & \textbf{DAM} & \textbf{RTM} \\
\hline \hline
Mean        & 3.1	& 39.1	& 36.0   \\\hline
Standard Deviation         & 87.7	& 45.6 &	84.5  \\\hline
Min           & -1488.6	& -61.6	& -262.3  \\\hline
Max             & 2276.2	& 2374.4	& 1545.4 \\\hline
25th Percentile  & -0.16	& 25.0	& 19.6 \\\hline
50th Percentile  & 6.6	& 33.4	& 27.1  \\\hline
75th Percentile & 17.5	& 46.5 &36.7\\\hline
\end{tabular}
\vspace{-0.21cm}
\label{tab:statistics}
\end{table}
It is worth pointing out that the standard deviation for the DAM was almost the half of that value for the RTM, which means that the values tend to be closer to the mean in case of DAM and prices do not fluctuate as much as the price fluctuations in case of RTM. The 25th percentile of all gap prices are less than -0.16 \$/MWh which means that almost one quarter of the direct gap prices are negative.

\par
In Tables \ref{tab:dam} and \ref{tab:rtm}, the evaluation of DAM price and RTM price procured by the learning methods given in Section \ref{Learning_methods} is presented. All the mentioned algorithms performed significantly better in predicting DAM prices than in predicting RTM electricity prices. For instance, LSTM the best performing algorithm had MAE and RMSE 4.9 and 7.1 respectively, while, for the RTM, the same algorithm resulted in MAE of 21.2 and RMSE of 48. Consequently, the complexity of predicting the gap between these markets is bounded by the accuracy of the prediction of the RTM.
\begin{table}[h!]
\small \centering
\centering
\vspace{-0.21cm}
\caption{Prediction errors for DAM}\vspace{-0.2cm}
\begin{tabular}{lllll} \hline 
\textbf{Error Measure} & \textbf{LASSO} & \textbf{SVR} & \textbf{RF} & \textbf{LSTM} \\ \hline \hline
MAE                    & 9.6  & 11.7 & 5.1  & 4.9 \\\hline
RMSE                    & 13.3   & 33.7 & 7.9  & 7.1  \\\hline
nRMSE{[}\%{]}           & 7.4   & 39.8 & 4.4  & 4.2  \\\hline
Max Error             & 95.2   & 122.7 & 62  & 40  \\\hline
\end{tabular}
\vspace{-0.21cm}
\label{tab:dam}
\end{table}

\begin{table}[h!]
\small \centering
\centering
\vspace{-0.21cm}
\caption{Prediction errors for RTM}\vspace{-0.2cm}
\begin{tabular}{lllll} \hline 
\textbf{Error Measure} & \textbf{LASSO} & \textbf{SVR} & \textbf{RF} & \textbf{LSTM} \\ \hline \hline
MAE                    & 18.9  & 21.4 & 26.4  & 21.2 \\\hline
RMSE                    & 59   & 54 & 71.9  & 48  \\\hline
nRMSE{[}\%{]}           & 5.1   & 5.0 & 6.2 & 4.4  \\\hline
Max Error             & 1064   & 1060 & 1058  & 1040  \\\hline
\end{tabular}
\vspace{-0.21cm}
\label{tab:rtm}
\end{table}

\begin{table}[h!]
\small \centering
\centering
\vspace{-0.21cm}
\caption{Prediction errors for Gap}\vspace{-0.2cm}
\begin{tabular}{lllll} \hline 
\textbf{Error Measure} & \textbf{LASSO} & \textbf{SVR} & \textbf{RF} & \textbf{LSTM} \\ \hline \hline
MAE                    & 19.6  & 28.2 & 24.5  & 17.1 \\\hline
RMSE                    & 58.9 & 80.4  & 67.5  & 56.9  \\\hline
nRMSE{[}\%{]}           & 4.98 & 6.1  & 5.7  & 4.8  \\\hline
Max Error             & 1054.8   & 1051 & 1048  & 1046  \\\hline
\end{tabular}
\vspace{-0.21cm}
\label{tab:gap}
\end{table}

\begin{figure}[h]
\centering
\includegraphics[width=0.5\textwidth]{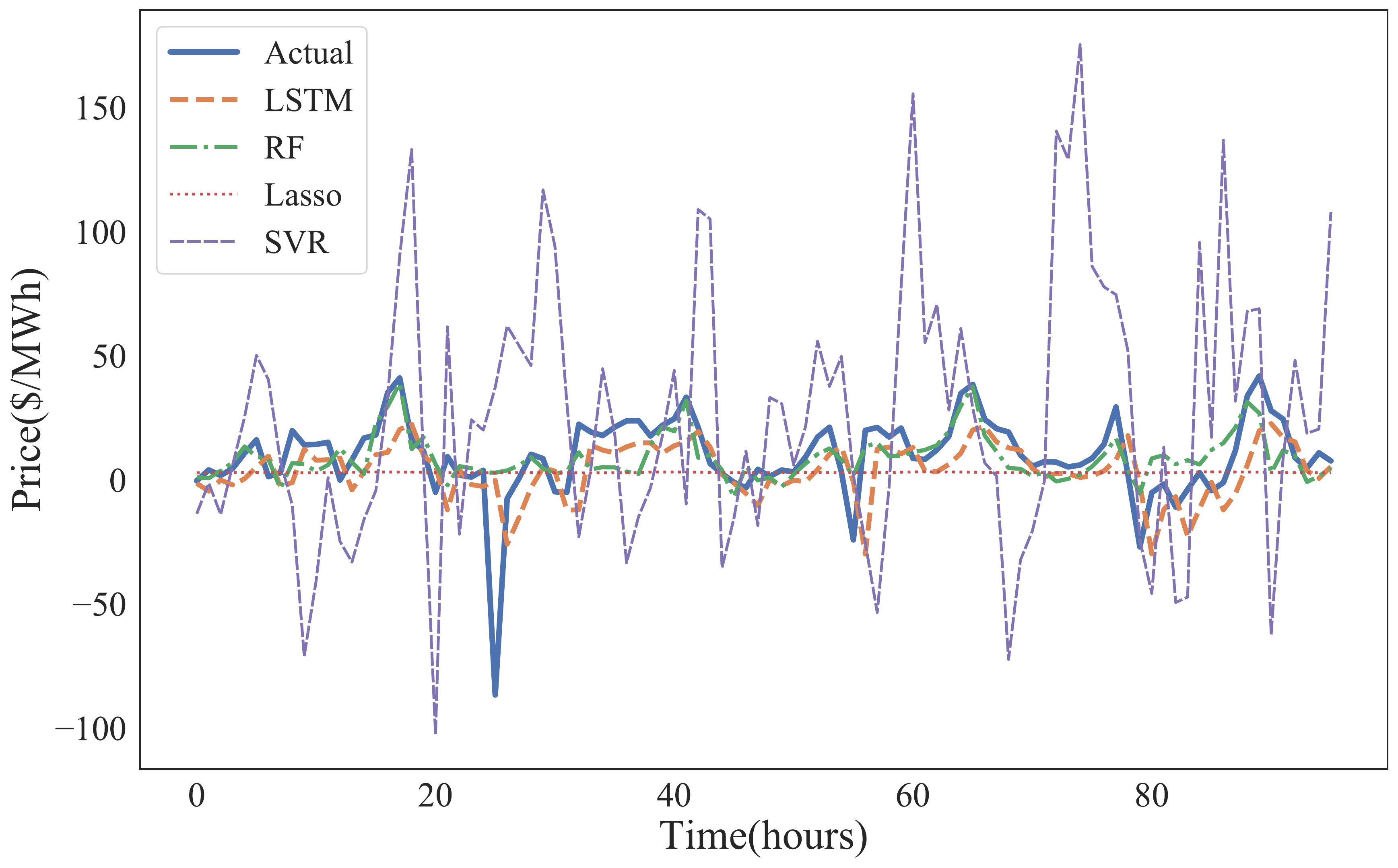}
\caption{The comparison of a gap forecast for the next 96 hours to the actual values of gap.}
\label{fig:results_gap}
\end{figure}

Lasso algorithm performs very poorly to predict any trend in gap forecast and outputs an almost constant predictions for all timestamps in the testing set. The range of the predictions is [1.84, 3.56].
The Lasso failed to capture any spikes in the price change and the maximum error between the prediction and the actual gap values were 1054.8. The poor performance of the Lasso algorithm can be explained by the fact that the Lasso is a linear algorithm and uses a linear function for prediction, while gap prediction should be non-linear mapping based on empirical evidence.
SVR with kernel function of radial basis performs better than Lasso, however, the results are still not accurate enough. Even though the mentioned algorithm does not output almost constant predictions as in the case of the Lasso algorithm, it still either overestimates or underestimates the price of the gap value as it is shown in the Fig. ~\ref{fig:results_gap}. 
Even though SVR showed some improvements in terms of predicting gap price spikes, it is not as good as Random Forest and LSTM algorithms. Random Forest algorithm had an MAE score of 24.5 and an RMSE score of 67.5 when predicting direct gap prices. Even though these metrics are slightly worse than the above-described algorithms, it can be observed from the Fig. ~\ref{fig:results_gap} that Random Forest is performing better than Lasso and SVR, but still not as good as LSTM. Even though random forest outputs almost correct values when predicting positive gap prices, it suffers to capture big negative price spike in gap values.
LSTM algorithm had the best performance in terms of error metrics as well as empirical evaluation based on the plot provided in Fig. ~\ref{fig:results_gap}. 
Table~\ref{tab:gap} shows that the LSTM algorithm procures lowest values of MAE, nRMSE, and Max Error values among all learning methods to predict gap price. In addition, LSTM also outperformed all the methods in predicting DAM and RTM electricity prices. \vspace{-0.35cm}
\subsection{Importance of Exogenous Information}
To illustrate the importance of the collected exogenous features, the LSTM model is leveraged without exogenous features to predict gap prices and the results are compared to those of the models described in Section ~\ref{Learning_methods}, using collected features. The LSTM model without collected exogenous features had the largest MAE score of $31.8$ out of all models that used collected features. RMSE score of the mentioned model turned out to be $62.15$, which is higher than the RMSE score of LSTM model using collected features. In addition, nRMSE=$5.2$ error metric is also worse than that of a LSTM model with collected features. Consequently, adding exogenous features such as weather conditions, and solar radiation significantly improved the accuracy of the gap prediciton.

\vspace{-0.35cm}
\section{Conclusions}
This paper leverages statistical machine learning algorithms and neural networks to predict the price gap between the DAM LMP and RTM LMP. Besides, several exogenous features like weather features including solar irradiance are collected and the impact of these features on predicting the DAM and RTM price gap is investigated. For this purpose, first, the data preparation is illustrated, where three distinct datasets are collected. Then LASSO, SVR, RF, and LSTM learning methods are described and various methods to evaluate the performance of the learning algorithms are presented.  The significance of features collected is illustrated and the benefits of directly predicting the gap as well as the probability distribution of the predicted prices are shown. In addition, the feature selection was done and externally collected features turned out to have an impact on the direct gap prediction. Furthermore, four different algorithms were compared to predict the gap prices and their performance was evaluated based on the error metrics mentioned  for an arbitrary chosen bus within California ISO.

\par


%





\ifCLASSOPTIONcaptionsoff
\fi




\bibliographystyle{IEEEtran}
\bibliography{bib/literature.bbl}
\end{document}